\documentclass[preprint,superscriptaddress,amsmath,floats]{revtex4}
\usepackage{txfonts}
\usepackage{amssymb}
\usepackage{graphicx}

\begin{document}

\title{Metal-Insulator Transition and Emergent Gapped Phase in the Surface-Doped 2D Semiconductor 2H-MoTe$_2$ }

\author{T. T. Han}
\affiliation{International Center for Quantum Materials, School of Physics, Peking University, Beijing 100871, China}

\author{L. Chen}
\affiliation{International Center for Quantum Materials, School of Physics, Peking University, Beijing 100871, China}

\author{C. Cai}
\affiliation{International Center for Quantum Materials, School of Physics, Peking University, Beijing 100871, China}

\author{Z. G. Wang}
\affiliation{International Center for Quantum Materials, School of Physics, Peking University, Beijing 100871, China}

\author{Y. D. Wang}
\affiliation{International Center for Quantum Materials, School of Physics, Peking University, Beijing 100871, China}

\author{Z. M. Xin}
\affiliation{International Center for Quantum Materials, School of Physics, Peking University, Beijing 100871, China}

\author{Y. Zhang}\email{yzhang85@pku.edu.cn}
\affiliation{International Center for Quantum Materials, School of Physics, Peking University, Beijing 100871, China}
\affiliation{Collaborative Innovation Center of Quantum Matter, Beijing 100871, China}

\date{\today}

\begin{abstract}

Artificially created two-dimensional (2D) interfaces or structures are ideal for seeking exotic phase transitions due to their highly tunable carrier density and interfacially enhanced many-body interactions. Here, we report the discovery of a metal-insulator transition (MIT) and an emergent gapped phase in the metal-semiconductor interface that is created in 2H-MoTe$_2$ via alkali-metal deposition. Using angle-resolved photoemission spectroscopy, we found that the electron-phonon coupling is strong at the interface as characterized by a clear observation of replica shake-off bands. Such strong electron-phonon coupling interplays with disorder scattering, leading to an Anderson localization of polarons which could explain the MIT. The domelike emergent gapped phase could then be attributed to a polaron extended state or phonon-mediated superconductivity. Our results demonstrate the capability of alkali-metal deposition as an effective method to enhance the many-body interactions in 2D semiconductors. The surface-doped 2H-MoTe$_2$ is a promising candidate for realizing polaronic insulator and high-$T_c$ superconductivity.

\end{abstract}

\pacs{74.25.Jb,74.70.Xa,79.60.-i}

\maketitle

Metal-insulator transitions (MITs) driven by many-body interactions attract great interests in condensed matter physics \cite{gap1, gap2, gap3, gap3a, gap3b, gap5, gap6, gap6a, gap7, gap8, gap11, gap15, gap16}. For example, in strongly correlated materials, a Mott insulator transition occurs when the hopping of electrons is prohibited by strong on-site Coulomb repulsions \cite{gap2}. In the materials with disorders or impurities, the disorder scattering of electrons leads to an Anderson localization, which drives a MIT \cite{gap3, gap3a}. In polaronic materials, due to strong electron-phonon coupling, moving electrons are dressed by lattice excitations, forming composed quasiparticles, polarons. A MIT occurs when polarons are self-trapped by disorder scattering \cite{gap5, gap6, gap6a} or electronic correlation \cite{gap7, gap8}. Besides the MITs themselves, exotic phases always emerge at the MIT transition point where quantum fluctuations arise from the competition among different degrees of freedom. For example in cuprates, high-$T_c$ superconductivity, charge-density-waves, and pseudogap phases all emerge near the MIT transition point \cite{gap11}.

Recently, the discoveries of MITs in LaAlO$_3$/SrTiO$_3$, twisted bilayer graphene, etc., ignited an intensive wave of research on seeking exotic MITs in the artificially created two-dimensional (2D) interfaces or structures \cite{gap15, gap16}.  Taking advantage of 2D materials such as simple structure, low dimensionality, and highly tunable carrier density, the 2D interfaces or structures with interfacially enhanced many-body interactions are ideal for simulating strongly correlated phenomena and seeking new exotic phases. 2H-MoX$_2$ (X = S, Se, and Te) are graphenelike 2D semiconductors, which can be easily exfoliated from bulk to monolayer \cite{gap18, gap19, gap20}. Electron-phonon coupling plays an important role in electron-doped 2H-MoX$_2$ due to its multivalley characteristic of the conduction bands. Superconductivity emerges in ionic-liquid-gated 2H-MoX$_2$ \cite{gap21, gap22, gap23}, which is reckoned to originate from strong electron-phonon coupling \cite{gap24, gap25, gap26, gap27}. In alkali-metal intercalated 2H-MoX$_2$, strong electron-phonon coupling results in a phonon softening which drives a 2H-to-1T structural transition \cite{gap28, gap29}.

Here, we report the discovery of a distinctive MIT in the surface-doped 2H-MoTe$_2$ via angle-resolved photoemission spectroscopy (ARPES) and alkali-metal deposition. Unlike the electrostatic gating or alkali-metal intercalation that donates electrons in the entire bulk materials, the surface deposition of alkali-metal only dopes electrons in the topmost layer, which naturally creates a 2D surface-metal/bulk-semiconductor interface. We show that the MIT that occurs at the interface could be attributed to an Anderson localization of polarons which originates from an interfacially enhanced electron-phonon coupling and weak disorder scattering. At the MIT transition point, a domelike gapped phase emerges with a gap closing temperature as high as 65~K. Its origin is discussed, considering a polaronic extended state and phonon-mediated superconductivity.

High-quality single crystals of 2H-MoTe$_2$  were synthesized using the chemical vapour transport method \cite{gap32}. ARPES measurements were performed at Peking University using a DA30L analyzer and a helium discharging lamp. The photon energy is 21.2~eV. The overall energy resolution was $\sim$8~meV and the angular resolution was $\sim$0.3$^\circ$. Core-level spectroscopy and low energy electron diffraction (LEED) were measured at the BL03U beam line of Shanghai Synchrotron Radiation Facility (SSRF). The crystals were cleaved $in$ $situ$ and measured in vacuum with a base pressure better than 6 $\times$10$^{-11}$~mbar. The alkali-metals (Li, Rb) were deposited $in$ $situ$ using an alkali-metal dispenser. We repeated the same deposition procedure several times. The deposition sequence is denoted using Dn for Rb doping and LDn for Li doping (n is the doping times). The total alkali-metal coverage is estimated to be $\sim$0.2~ML for Rb doping and $\sim$0.5~ML for Li doping. The alkali-metal deposition was conducted at 25~K to avoid any alkali-metal intercalation \cite{supp}. The Fermi energy ($E_F$) was determined using a gold reference. No surface charging effect is observed in the entire doping range \cite{supp}. A temperature dependent experiment was conducted upon cooling to avoid any alkali-metal desorption. We checked the spectral changing using temperature cycles. The alkali-metal desorption is negligible.

\begin{figure*}[t]
\includegraphics[width=15.8cm]{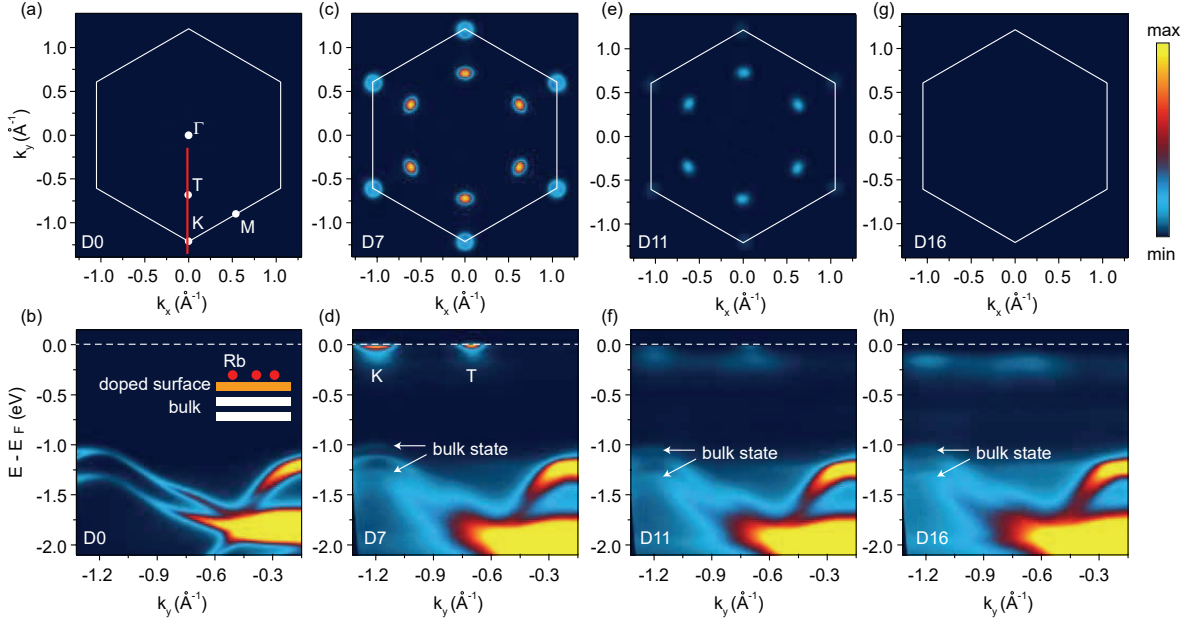}
\caption{Doping evolutions of Fermi surface and band structure in the surface-doped 2H-MoTe$_2$. (a) Fermi surface mapping taken in the as-grown sample. (b) The energy-momentum cut taken along the $\Gamma$-K direction in the as-grown sample. (c) and (d), (e) and (f), (g) and (h) are the same as (a) and (b) but taken in D7, D11 and D16 sample. Dn (n is the doping times) denotes the deposition sequence.}\label{f1}
\end{figure*}

As shown in Fig.~\ref{f1}, the as-grown (D0) bulk 2H-MoTe$_2$ is a semiconductor with a band gap of $\sim$1.1~eV [Figs.~\ref{f1}(a) and \ref{f1}(b)]. Two valence bands are clearly identified at $\sim$1.1 and $\sim$1.3~eV at the K point. The alkali-metal deposition causes a band bending at the sample surface, resulting in a charge transfer from the bottom layer to the topmost layer \cite{gap30, gap31, gap33}. Such charge transfer creates an energy separation between the surface and bulk electronic states. The probing depth of ARPES is normally larger than one layer. Therefore, we could detect both states as shown in Fig.~\ref{f1}(d). The valence bands at the K point split into two groups. One set of bands shifts slightly towards $E_F$, indicating a loss of electrons, which could be attributed to the semiconducting bulk state. The other set of bands shifts towards higher binding energy, which could be attributed to the electron-doped surface state. With electron doping, the surface state becomes metallic as electrons fill into the conduction band bottoms. Two tiny electron pockets emerge at the K and T points, which is consistent with the band calculations \cite{gap19, gap33, gapband}. The separation between surface and bulk electronic states confirms that the alkali-metal deposition creates a metal-semiconductor interface between the surface and bulk layers in 2H-MoTe$_2$. Intriguingly, with further deposition, the quasiparticle spectral weight transfers to higher binding energy [Figs.~\ref{f1}(d) and \ref{f1}(f)]. The system enters an insulating state in D16 with no residual spectral weight at $E_F$ [Figs.~\ref{f1}(g) and \ref{f1}(h)].

\begin{figure*}[t]
\includegraphics[width=15.8cm]{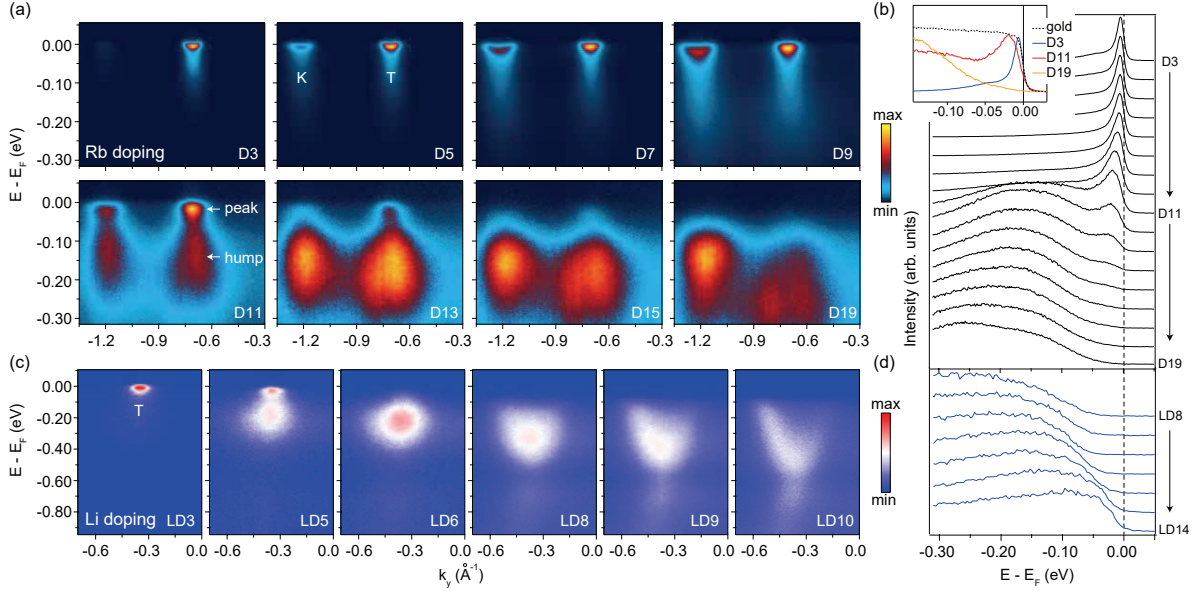}
\caption{Doping dependence of the conduction bands in the surface-doped 2H-MoTe$_2$. (a) Doping dependence of the energy-momentum cut taken along the $\Gamma$-K direction in Rb-doped samples.  (b) Doping dependence of the energy distribution curve (EDC) taken at the Fermi crossing ($k_F$) of the T electron band in Rb-doped samples. Inset panel shows the EDCs taken in D3, D11, D19 and reference gold. (c) and (d) are the same as (a) and (b), but taken in Li-doped samples crossing the T point along the M-K direction. }\label{f2}
\end{figure*}

The detailed evolution of the conduction bands is shown in Fig.~\ref{f2}. The system is metallic in the dilute-doped region (D3-D9) as characterized by the well-defined quasiparticle peaks and clear Fermi cutoff [Fig.~\ref{f2}(b)]. From D10, the quasiparticle peak becomes ill defined. Its spectral weight transfers to higher binding energy, forming an incoherent hump at around -0.15~eV. Eventually, the quasiparticle peak vanishes in D16. In the Li-doped sample, we manage to extend the doping range to a more heavily doped region \cite{supp}. The vanishing of quasi-particles occurs now at LD6 [Fig.~\ref{f2}(c)]. With further doping, the electron band dispersion recovers gradually. The spectrum sharpens and its leading edge shifts toward $E_F$ [Fig.~\ref{f2}(d)]. Note that, an additional feature emerges at $\sim$-0.7~eV in heavily doped sample which could be attributed to an extrinsic state originated from alkali-metal intercalation or alkali-metal cluster formation.

The reentrance of electron band dispersion as well as our LEED and core-level characterizations \cite{supp} clearly exclude the disorder scattering of photoelectrons or the surface degradation as a cause of spectral broadening, which suggests that the spectral broadening and resharpening originate from a localization and delocalization of quasiparticles. The vanishing of spectral weight at $E_F$ and the formation of an incoherent hump are all consistent with the APRES characterizations of quasiparticle localization \cite{gap33a,gap33b}, indicating that the system enters an insulating state in the intermediately doped region. With further doping, the reentrance of band dispersion reflects a delocalization of quasiparticles. The metallic property of the system recovers. We note that there is no clean Fermi cutoff in LD14, which suggests that the system is still in a bad-metallic state where a gap opens at $E_F$.

\begin{figure}[t]
\includegraphics[width=8.6cm]{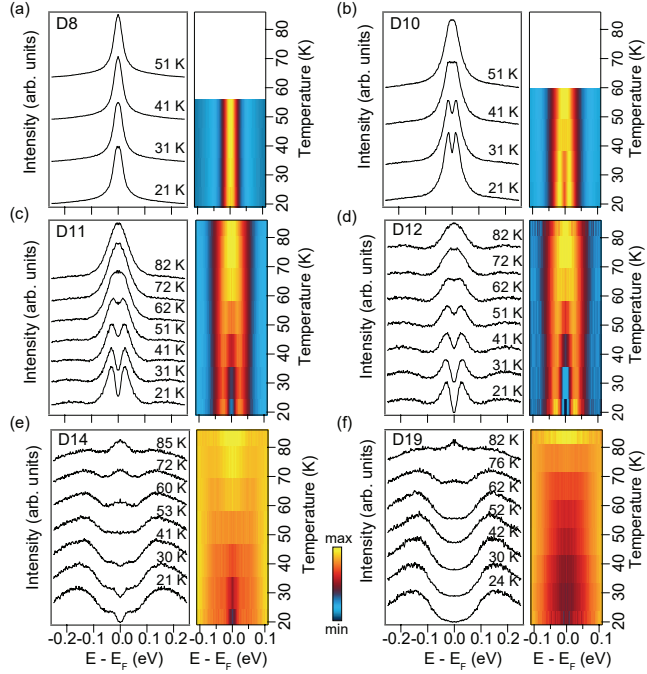}
\caption{Temperature dependence of the symmetrized EDCs in the surface-doped 2H-MoTe$_2$. (a) The temperature dependence of the symmetrized EDCs taken at the $k_F$ of the T electron band in D8. Right panel is the merged image of the symmetrized EDCs. (b)-(f) are the same as (a) but taken in D10, D11, D12, D14 and D19, respectively. }\label{f3}
\end{figure}

Besides the large insulating gap, the quasiparticle leading edge shifts below $E_F$ in D11 [Fig.~\ref{f2}(b)], indicating a quasiparticle gap opening. It is then intriguing to study how this energy gap evolves with temperature and electron doping. Although the spectral function is not necessarily particle-hole symmetric, we could determine the gap closing temperature using the EDC symmetrization \cite{supp,gap34}. As shown in Fig.~\ref{f3}, upon cooling, a peak to dip transition occurs at $E_F$ indicating an energy gap opening. The gap closing temperature first increases from 40 to 65~K when going from D10 to D12. It then decreases to around 40~K with further electron doping. In D19, there is no energy gap opening. A continuous suppression of quasiparticle spectral weight occurs upon cooling.

\begin{figure*}[t]
\includegraphics[width=15.8cm]{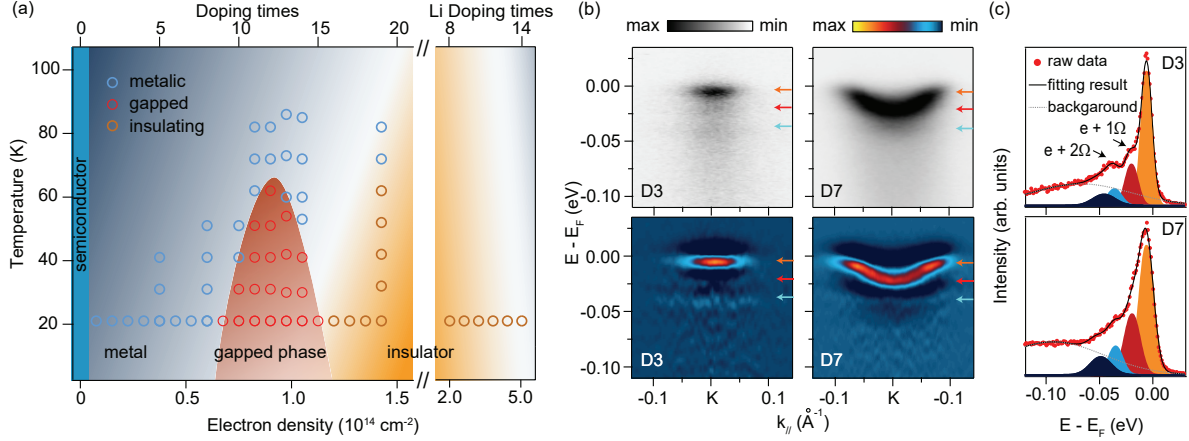}
\caption{Phase diagram and strong electron-phonon coupling in the surface-doped 2H-MoTe$_2$. (a) Phase diagram of the surface-doped 2H-MoTe$_2$. We calculated the carrier density by counting the Fermi surface volume \cite{supp}. The different colors of the data points illustrate the different line shapes of ARPES spectra as shown in Fig.~\ref{f2}(b). The metallic phase is referred to as a phase where a quasiparticle peak exists and there is no energy gap at $E_F$. The gapped phase is referred to as a phase where the quasiparticle peak exists but opens an energy gap at $E_F$. The insulating phase is referred to as a phase where no quasiparticle peak is present near $E_F$. (b)  Energy momentum cuts taken along the $\Gamma$-K direction in D3 and D7 samples. Upper and lower panels show the raw and second derivative images, respectively. (c) The peak fitting of the EDCs taken at the $k_F$ of the data in (b). }\label{f4}
\end{figure*}

We could then delineate the electronic phase diagram of the surface-doped 2H-MoTe$_2$ [Fig.~\ref{f4}(a)]. The obtained phase diagram resembles the phase diagrams of cuprates, LaAlO$_3$/SrTiO$_3$, and twisted bilayer graphene \cite{gap11,gap15,gap16}, where the localization and itinerancy of electrons compete with each other, resulting in a MIT and a domelike emergent phase at the MIT transition point. To understand why quasiparticles localize in the surface-doped 2H-MoTe$_2$, we consider both Mott localization and Anderson localization. Here, unlike cuprates and twisted bilayer graphene, the band filling is estimated to be below $\sim$0.2 electron per unit cell far from half-filling \cite{supp}, which excludes the Mott scenario. On the other hand, the alkali-metal adatoms distribute randomly on the sample surface, acting as random scattering centers, which favors the Anderson localization. However, a typical Anderson localization of electrons broadens the photoemission spectra and suppresses the spectral weight at $E_F$, but would not renormalize the entire band structure \cite{gap3, gap3a, gap3b, gap5, gap6}. Here, we observed a huge energy renormalization of spectra as characterized by the peak-dip-hump structure, where coherent spectra no longer peak at the center of mass of the ARPES spectrum. This cannot be explained by a pure Anderson localization of electrons. Moreover, there is no reported evidence to show that alkali-metal adatoms are strong scattering or trapping centers. Pure disorder scattering cannot explain the large energy scale of the spectral weight transfer observed here.

We note that the electron-phonon coupling is strong in bulk-doped 2H-MoX$_2$ \cite{gap24, gap25, gap26, gap27}. Here, in the surface-doped 2H-MoTe$_2$, the electron-phonon coupling could be further enhanced at the 2D metal/semiconductor interface \cite{gap39}. The low dimensionality and low free carrier density lead to a poor screening of Coulomb interactions. The dielectric constant is also strongly anisotropic at the surface-bulk interface due to the electrostatic potential generated by alkali-metal adatoms. We took high statistical scan near the K point in the dilute-doped samples. From the raw and second derivative images [Fig.~\ref{f4}(b)], the replica bands are clearly resolved. Similar behavior has been observed in 1ML FeSe/SrTiO$_3$, surface-doped 2H-MoS$_2$, SrTiO$_3$ and LaAlO$_3$/SrTiO$_3$ \cite{gap39, gap40, gap37, gap38, gap41}, and were attributed to the shake-off excitations involving a bosonic mode and coupled electrons. Here, we found that the energy separation between each replica band is around 17~meV, which is consistent with the mode energies of the $A_{1g}$ and $E_{1g}$ optical phonons in MoTe$_2$ \cite{gap43}. According to a simple theoretical model of electron-phonon coupling \cite{supp, gap39, gap37, gap38, gap35, gap36}, the electron-phonon interaction constant ($a_c$) could be estimated from the intensity ratios of the replica bands ($I_n$) to the main band ($I_0$) using the Poisson distribution $I_n$/$I_0$ = $a_c^{2n}$/$n!$. The $I_1$/$I_0$ and $I_2$/$I_0$ are around 0.3 and 0.12 in D3, but increase to 0.47 and 0.22 in D7. Correspondingly, $a_c$ increases from 0.5-0.6 in D3 to 0.7-0.8 in D7.

The observation of strong electron-phonon coupling is important in explaining the MIT in the surface-doped 2H-MoTe$_2$. It is known that Anderson localization could be strongly influenced by electron-phonon coupling \cite{gap5,gap6,gap6a}. In general, electron-phonon coupling decreases the itinerancy of electrons which helps Anderson localization. When electron-phonon coupling is strong, an Anderson localization of polarons occurs in the presence of weak disorder scattering \cite{gap5,gap6,gap6a}. A carrier is confined with a lattice distortion and then self-traps due to disorder scattering. Such localization process involves not only the disorder scattering potential but also the binding energy of polarons, which could explain the large energy scale of the spectral weight transfer observed here. It also explains the reentrance of the electron band dispersion. With alkali-metal deposition, the increments of carrier density and bandwidth suppress the formation of polarons. As a result, the itineracy of quasiparticles recovers. The band structure of the heavily doped sample is now consistent with the calculated band structure after a rigid-band shift \cite{supp}, indicating that the system is better described using a single-electron description. While the spectral weight suppresses at $E_F$, the entire band structure is not renormalized, which is more consistent with an Anderson localization of electrons considering no electron-phonon coupling \cite{gap3,gap3a,gap3b}.

Finally, we discuss the origin of the emergent gapped phase. On one hand, a polaronic extended state may exist at the MIT transition boundary of a polaron-localized system \cite{gap6, gap9,gap10}. Such a polaronic extended state exhibits pseudogap behaviors, which may explain the emergent gapped phase. However, the shape of the polaronic extended state is normally determined by the MIT boundary \cite{gap6, gap9, gap10} which cannot explain the domelike shape of the emergent gapped phase. Moreover, the small quasiparticle gap and the large insulating gap are two different energy scales. Instead of a continuous gap closing, spectral weight transfers between them, which contradicts to the spectral evolution predicted in a polaronic extended state \cite{gap6, gap9, gap10}. On the other hand, superconductivity has been observed in bulk-doped 2H-MoX$_2$ (X = S, Se, and Te) \cite{gap21, gap22, gap23}. The T$_c$ is as high as 11~K. Here, the superconducting pairing strength could be strongly enhanced at the metal/semiconductor interface, which may explain the high gap closing temperature of the emergent gapped phase. We note that, the spectral characterizations of a superconductor, such as the back bending of band and sharp Bogliubov quasiparticle peaks, were not observed, which could be due to our high experimental temperature and the low dimensionality of the system.

In summary, our results highlight the rich phase diagram of the surface-doped 2H-MoTe$_2$, which originates from disorder scattering and an interfacially enhanced electron-phonon coupling. It is a promising system for seeking polaronic insulators and possible superconductivity. Further studies using transport and scanning tunneling spectroscopy could be very intriguing. More importantly, our observation demonstrates the capacity of surface alkali-metal deposition as a practical method that creates 2D surface-bulk interfaces with boosted electron-phonon and electron-electron interactions. This makes the surface-doped 2D materials, especially surface-doped 2D semiconductors, as a new category of materials to search for strongly correlated phenomena and exotic electronic phases.

This work is supported by the National Natural Science Foundation of China (No.~11888101), the National Key Research and Development Program of China (No.~2018YFA0305602 and No.~2016YFA0301003), and the National Natural Science Foundation of China (No.~91421107 and No.~11574004). The work at SSRF is supported by ME2 project under Contract No. 11227901 from National Natural Science Foundation of China.

\end{document}